\documentclass[preprint,showpacs,amsmath,amsfonts,amssymb,aps,noshowkeys]
              {revtex4}
\usepackage{latexsym,makeidx,ifthen,calc,longtable,graphics}
\usepackage{bm}

\begin{document}

\title{The Universe as an Eigenstate:\protect\\
Spacetime Paths and Decoherence}

\thanks{Presented to the 2006 Conference of the International
Association for Relativistic Dynamics} 

\preprint{Version 1.4 (ePrint v3)}

\author{Ed Seidewitz}
\email{seidewitz@mailaps.org}
\affiliation{14000 Gulliver's Trail, Bowie MD 20720 USA}

\date{18 March 2007}


\begin{abstract}
    This paper describes how the entire universe might be considered
    an eigenstate determined by classical limiting conditions within
    it. This description is in the context of an approach in which the
    path of each relativistic particle in spacetime represents a
    \emph{fine-grained} history for that particle, and a path integral
    represents a \emph{coarse-grained} history as a superposition of
    paths meeting some criteria. Since spacetime paths are
    parametrized by an invariant parameter, not time, histories based
    on such paths do not evolve in time but are rather histories of
    all spacetime. Measurements can then be represented by orthogonal
    states that correlate with specific points in such coarse-grained
    histories, causing them to decohere, allowing a consistent
    probability interpretation. This conception is applied here to the
    analysis of the two slit experiment, scattering and, ultimately,
    the universe as a whole. The decoherence of \emph{cosmological
    states} of the universe then provides the eigenstates from which
    our ``real'' universe can be selected by the measurements carried
    out within it.
\end{abstract}

\maketitle


\section{Introduction} \label{sect:intro}

Before the ascendency of quantum field theory, Stueckelberg proposed
an approach to relativistic quantum field theory based on the
conception of particle paths in spacetime, parameterized by an
invariant fifth parameter \cite{stueckelberg41, stueckelberg42}.
Feynman later considered this idea as the basis for relativistic path
integrals (see the appendices to \refcite{feynman50, feynman51}), a 
conception which seems to have informed his early work on quantum 
field theory (though it is not much apparent in his later work).

Since then, a number of authors have further developed the theory of
parameterized relativistic quantum physics (see \refcite{fanchi93} and
references therein), though not necessarily using a path integral
approach. However, relativistic path integrals in particular have a
natural interpretation in terms of consistent or decoherent histories
\cite{griffiths84, omnes88, gellmann90}. In this interpretation, the
path of a particle in spacetime is considered a \emph{fine-grained}
history. A path integral then represents a \emph{coarse-grained}
history as a superposition of paths meeting some criteria. When the
criteria are properly chosen, the states for these coarse grained
histories do not interfere---that is, they are \emph{decoherent}
\cite{hartle95}.

Since decoherent histories do not interfere, they can be assigned
classical probabilities. Further adopting a ``many worlds''
interpretation \cite{everett57}, these histories can be considered to
be alternate ``branches'' in the history of the universe, with
associated probabilities for each of the branches to ``occur''. (For
an informal introduction to the ideas of decoherence and emergent
classicality, see \refcite{halliwell04}. For a more extensive survey
see \refcite{halliwell03}.)

Relativistic path integrals have also proved useful in the study of
quantum gravity and quantum cosmology, because the time coordinate is
treated similarly to the space coordinates, rather than as an
evolution parameter (see, for example, \refcite{teitelboim82,
hartle95}). In quantum cosmological models, the total Hamiltonian
annihilates the ``wave function of the universe'', rather than
determining the time evolution of the system. The question is how to
extract physical predictions from such a wave function.

Inspired by this, Halliwell and Thorwart recently published a paper
with the engaging title ``Life in an energy eigenstate''
\cite{halliwell02} in which they consider the internal dynamics of a
simple particle system in an energy eigenstate. In the present paper,
I would like to take this idea a bit farther, and describe how the
\emph{entire universe} might be considered to be in an eigenstate
determined by classical limiting conditions within it. In effect, such
an eigenstate is a selection of a specific coarse-grained branch as
``the'' history of the universe.

Pursuing this idea requires a formalism that allows coarse-grained
histories to be expressed as quantum states. I will use the spacetime
path formalism proposed in \refcite{seidewitz06a}. For completeness,
\sect{sect:formalism} summarizes the development of this formalism. A
particularly important result from this work is that the
coarse-grained histories of free particles with fixed 3-momentum
become on-shell and decoherent in the infinite time limit.

\Sect{sect:decoherence} then discusses decoherence in the context of
the spacetime path formalism. \Sect{sect:decoherence:2slit} applies
the formalism to the analysis of the familiar scenario of the two slit
experiment. \Sect{sect:decoherence:scattering} extends the approach to
consideration of a scattering process that takes place in a finite
region of spacetime. Finally, taking this analysis of scattering as a
paradigm, \sect{sect:decoherence:probabilities} considers the relation
of probabilities to measured relative frequencies and
\sect{sect:decoherence:cosmo} presents a heuristic discussion of the
decoherence of \emph{cosmological states} of the entire universe.

Throughout, I will use a spacetime metric signature of $(-+++)$ and
take $\hbar = c = 1$.

\section{Spacetime Paths} \label{sect:formalism}

This section summarizes the spacetime path formalism I will use in 
the following sections. For further details on the development of 
this formalism, see \refcite{seidewitz06a}.

\subsection{Position States} \label{sect:formalism:position}

A \emph{spacetime path} is specified by four functions $\qmul$, for
$\mu = 0, 1, 2, 3$, of a \emph{path parameter} $\lambda$. Note that
such a path is not constrained to be timelike or even to maintain any
particular direction in time. The only requirement is that it must be
continuous. And, while there is no \emph{a priori} requirement for the
paths to be differentiable, we can, as usual, treat them as
differentiable within the context of a path integral (see the
discussion in \refcite{seidewitz06a}.)

It is well known that a spacetime path integral of the form
\begin{equation} \label{eqn:A1}
    \prop = \eta \int_{\lambdaz}^{\infty} \dif \lambda_{1}\, \intDfour q\,
            \delta^{4}(q(\lambda_{1}) - x) \delta^{4}(q(\lambdaz) - \xz)
            \exp\left( 
                \mi \int^{\lambda_{1}}_{\lambdaz} \dl L(\qdotsq(\lambda))
            \right) \,,
\end{equation}
for an appropriate normalization constant $\eta$ and the Lagrangian
function
\begin{equation*}
    L(\qdotsq) = \frac{1}{4}\qdotsq - m^{2} \,,
\end{equation*}
gives the free-particle Feynman propagator \cite{feynman50,
teitelboim82, halliwell01b, seidewitz06a}. In the path integral above,
the notation $\Dfour q$ indicates that the integral is over the four
functions $\qmul$ and the delta functions constrain the starting and
ending points of the paths integrated over.

Consider, however, that \eqn{eqn:A1} can be written
\begin{equation*}
    \prop = \int_{\lambdaz}^{\infty} \dif \lambda_{1}\, 
            \kersym(x-\xz; \lambda_{1}-\lambdaz) \,,
\end{equation*}
where
\begin{equation} \label{eqn:A2}
    \kersym(x-\xz; \lambda_{1}-\lambdaz) \equiv
        \eta \intDfour q\,
            \delta^{4}(q(\lambda_{1}) - x) \delta^{4}(q(\lambdaz) - \xz)
            \exp\left( 
                \mi \int^{\lambda_{1}}_{\lambdaz} \dl L(\qdotsq(\lambda))
            \right)
\end{equation}
now has a similar path integral form as the usual non-relativistic
\emph{propagation kernel} \cite{feynman48, feynman65}, except with
paths parametrized by $\lambda$ rather than time. We can, therefore,
use the relativistic kernel of \eqn{eqn:A2} to define a parametrized
probability amplitude function in a similar fashion to the
non-relativistic case:
\begin{equation} \label{eqn:A3}
    \psixl = \intfour \xz\, \kerneld \psixlz \,.
\end{equation}
These wave functions are just the parametrized probability amplitude
functions defined by Stueckelberg \cite{stueckelberg41}. In this
sense, the $\psixl$ represent the probability amplitude for a particle
to reach position $x$ at the point along its path with parameter value
$\lambda$.

The path integral in \eqn{eqn:A2} can be evaluated to give
\cite{teitelboim82, seidewitz06a}
\begin{equation*}
    \kerneld = (2\pi)^{-4} \intfour p \, \me^{\mi p\cdot(x - \xz)}
               \me^{-\mi (\lambda-\lambdaz)(p^{2} + m^{2})} \,.
\end{equation*}
Inserting this into \eqn{eqn:A3}, we see that $\psixl$ satisfies the
\emph{Stuekelberg-Schr\"odinger equation}
\begin{equation*}
    -\mi \pderiv{}{\lambda} \psixl 
        = \left( 
            \frac{\partial^{2}}{\partial x^{2}} - m^{2}
          \right) \psixl \,.
\end{equation*}
Note that this equation is based on the relativistic Hamiltonian
$p^{2}+m^{2}$, and therefore includes the mass term $m^{2}$. This is
in contrast to most previous authors \cite{feynman50, horwitz73,
fanchi78}, who used a Hamiltonian of the form $p^{2}/(2m)$, by analogy
with non-relativistic mechanics.

The relativistic propagation kernel can also be given a conjugate form
as a superposition of particle mass states. For $T > 0$,
\begin{equation} \label{eqn:A4}
    \begin{split}
        \theta(T)\kersym(x-\xz;T)
            &= \me^{-\mi T m^{2}} 
               \intfour p\, \me^{\mi p\cdot(x-\xz)}
               \int_{0}^{\infty} \dif T'\, \me^{-\mi T' p^{2}}
               \delta(T'-T) \\
            &= (2\pi)^{-1}\me^{-\mi T m^{2}}
               \intfour p\, \me^{\mi p\cdot(x-\xz)}
               \int_{0}^{\infty} \dif T'\, \me^{-\mi T' p^{2}}
               \int \dif m'^{2}\,
               \me^{-\mi(T'- T)m'^{2}} \\
            &= (2\pi)^{-1}\me^{-\mi T m^{2}}
               \int \dif m'^{2}\, \me^{\mi T m'^{2}}
               \propsym(x-\xz;m'^{2}) \,,
    \end{split}
\end{equation}
where
\begin{equation*}
    \propsym(x-\xz;m'^{2}) 
        \equiv \int_{0}^{\infty} \dif T'\,
                   \intfour p\, \me^{\mi p\cdot(x-\xz)}
                       \me^{-\mi T'(p^{2}+m'^{2})}
        =      -\mi(2\pi)^{-4}\intfour p\, 
                   \frac{\me^{\mi p\cdot(x-\xz)}}
                        {p^{2}+m'^{2}-\mi\varepsilon} \,.
\end{equation*}
Except for the extra phase factor $\exp(-\mi T m^{2})$, this form for
$\kersym(x-\xz;T)$ is essentially that of the retarded Green's
function derived by Land and Horwitz for parametrized quantum field
theory \cite{land91,frastai95} as a superposition of propagators for
different mass states (see also \refcite{enatsu63, enatsu86}).

The value $T$ in $\kersym(x-\xz;T)$ can be thought of as fixing a
specific \emph{intrinsic length} for the paths being integrated over
in \eqn{eqn:A2}. The full propagator then results from a regular
integration over all possible intrinsic path lengths:
\begin{equation*}
    \prop = \int_{0}^{\infty} \dif T\, \kersym(x-\xz; T) \,.
\end{equation*}
As a result of the phase factor $\exp(-\mi T m^{2})$ in \eqn{eqn:A4},
the integration over $T$ effectively acts as a Fourier transform,
resulting in the Feynman propagator with mass sharply defined at $m$,
$\prop = \propsym(x-\xz;m)$.

The functions defined in \eqn{eqn:A3} form a Hilbert space over four
dimensional spacetime, parameterized by $\lambda$, in the same way
that traditional non-relativistic wave functions form a Hilbert space
over three dimensional space, parameterized by time. We can therefore
define a consistent family of \emph{position state} bases $\ketxl$,
such that
\begin{equation} \label{eqn:A5}
    \psixl = \innerxlpsi \,,
\end{equation}
given a single Hilbert space state vector $\ketpsi$. These position
states are normalized such that
\begin{equation*}
    \inner{x'; \lambda}{x; \lambda} = \delta^{4}(x' - x) \,.
\end{equation*}
for each value of $\lambda$. Further, it follows from \eqns{eqn:A3} and
\eqref{eqn:A5} that
\begin{equation} \label{eqn:A6}
    \kerneld = \innerxlxlz \,.
\end{equation}
Thus, $\kerneld$ effectively defines a unitary transformation between 
the various Hilbert space bases $\ketxl$, indexed by the parameter
$\lambda$.
               
The overall state for propagation from $\xz$ to $x$ is given by the
superposition of the states for paths of all intrinsic lengths. If we
fix $\qmulz = \xmu_{0}$, then $\ketxl$ already includes all paths of
length $\lambda - \lambdaz$. Therefore, the overall state $\ketx$ for
the particle to arrive at $x$ should be given by the superposition of
the states $\ketxl$ for all $\lambda > \lambdaz$:
\begin{equation*} 
    \ketx \equiv \int_{\lambdaz}^{\infty} \dl\, \ketxl \,.
\end{equation*}
Then, using \eqn{eqn:A6},
\begin{equation*}
    \innerxxlz
           = \int_{\lambdaz}^{\infty} \dl\, \kerneld 
           = \int_{0}^{\infty} \dl\, \kersym(x-\xz; \lambda)
           = \prop \,.
\end{equation*}

\subsection{On-Shell States} \label{sect:formalism:on-shell}

The position states defined in \sect{sect:formalism:position} make no
distinction based on the time-direction of propagation of particles.
Normally, particles are considered to propagate \emph{from} the past
\emph{to} the future. Therefore, we can define normal particle states
$\ketax$ such that
\begin{equation} \label{eqn:B0}
    \innerxaxlz = \thetaax \prop \,,
\end{equation}
On the other hand, \emph{antiparticles} may be considered to propagate
from the \emph{future} into the \emph{past} \cite{stueckelberg41,
stueckelberg42, feynman49}. Therefore, antiparticle states $\ketrx$
are such that
\begin{equation} \label{eqn:B0a}
    \innerxrxlz = \thetarx \prop \,.
\end{equation}

Note that the particle/antiparticle distinction proposed here is
subtly different than that originally proposed by Stueckelberg
\cite{stueckelberg41, stueckelberg42}. Stueckelberg considered the
possibility that a single particle path might undergo a dynamical
interaction that could change the time direction of its propagation,
corresponding to what seemed to be a particle creation or annihilation
event when viewed in a time-advancing direction. In contrast, the
definitions of particle and antiparticle states given here depend only
on whether the \emph{end point} $x$ of the particle path is in the
future or past of its starting point $\xz$. Between these two points,
the path may move arbitrarily forward or backwards in time.

This division into particle and antiparticle paths depends, of course,
on the choice of a specific coordinate system in which to define the
time coordinate. However, if we take the time limit of the end point
of the path to infinity for particles and negative infinity for
antiparticles, then the particle/antiparticle distinction will be
coordinate system independent.

In taking this time limit, one cannot expect to hold the 3-position of
the path end point constant. However, for a free particle, it is
reasonable to take the particle \emph{3-momentum} as being fixed.
Therefore, consider the state of a particle or antiparticle with a
3-momentum $\threep$ at a certain time $t$:
\begin{equation*}
    \ketar{t,\threep}
        \equiv (2\pi)^{-3/2} \intthree x\, 
                \me^{\mi(\mp\Ep t + \threep\cdot\threex)} 
                \ketar{t,\threex} \,,
\end{equation*}
where $\Ep \equiv \sqrt{\threep^{2} + m^{2}}$. Now, as shown in
\refcite{seidewitz06a},
\begin{equation} \label{eqn:B1}
    \begin{split}
        \keta{t,\threep} &=
              (2\Ep)^{-1} \int_{-\infty}^{t} \dt_{0}\,
                     \ketalz{t_{0}, \threep} \quad \text{and} \\
        \ketr{t,\threep} &=
              (2\Ep)^{-1} \int_{t}^{+\infty} \dt_{0}\,
                     \ketrlz{t_{0}, \threep} \,,
    \end{split}
\end{equation}
where
\begin{equation*}
    \ketarlz{t, \threep}
        \equiv (2\pi)^{-3/2} \intthree x\, 
                \me^{\mi(\mp\Ep t + \threep\cdot\threex)} 
                \ketlz{t, \threex} \,.
\end{equation*}
Since
\begin{equation*}
    \inner{\advret{t', \threepp}; \lambdaz}
          {\advret{t, \threep}; \lambdaz} =
        \delta(t'-t) \delta^{3}(\threepp - \threep) \,,
\end{equation*}
we have, from \eqn{eqn:B1},
\begin{equation*}
    \inner{t, \advret{\threep}}{\advret{t_{0}, \threep_{0}{}}; \lambdaz} =
        (2\Ep)^{-1} \theta(\pm(t-t_{0})) 
                    \delta^{3}(\threep - \threep_{0}) \,.
\end{equation*}
If we now define the time limit particle and antiparticle states
\begin{equation} \label{eqn:B2}
    \ketarthreep \equiv \lim_{t \to \pm\infty} \ketartp \,,
\end{equation}
then
\begin{equation} \label{eqn:B3}
    \inner{\advret{\threep}}{\advret{t_{0}, \threep_{0}{}}; \lambdaz}
            = (2\Ep)^{-1} \delta^{3}(\threep - \threep_{0}) \,,
\end{equation}
for \emph{any} value of $t_{0}$. 

\Eqn{eqn:B3} is a natural introduction of an ``induced'' inner
product, in the sense of \cite{halliwell01b, hartle97}. To see how
this induced inner product may be used, consider, the two
Hilbert-space subspaces spanned by the normal particle states
$\ketalz{t, \threep}$ and the antiparticle states $\ketrlz{t,
\threep}$, for each time $t$. States in these subspaces have the form
\begin{equation*}
    \ketarlz{t,\psi} 
        = \intthree p\, \psi(\threep) \ketarlz{t,\threep} \,,
\end{equation*}
for any square-integrable function $\psi(\threep)$, with
\begin{equation*}
    \psi(\threep) = 
        (2\Ep)\inner{\advret{\threep}}{t,\advret{\psi}} \,.
\end{equation*}
Similarly, consider the dual subspaces spanned by the bra states 
$\braa{\threep}$ and $\brar{\threep}$, such that
\begin{equation*}
    \braar{\psi} 
        \equiv \intthree p\, \psi(\threep)^{*} \braar{\threep}
\end{equation*}
and
\begin{equation} \label{eqn:B3a}
    \psi(\threep)^{*} 
        = \inner{\advret{\psi}}{t,\advret{\threep}}(2\Ep) \,.
\end{equation}
As a result of \eqn{eqn:B3}, we get the traditional inner product
\begin{equation} \label{eqn:B4}
    (\psi', \psi)
        \equiv \inner{\advret{\psi'}}{t,\advret{\psi}}
        = \int \frac{\dif^{3} p}{2\Ep} \psi'(\threep)^{*} \psi(\threep) \,.
\end{equation}

With the inner product given by \eqn{eqn:B4}, the spaces of the
$\ketar{t,\psi}$ can be considered ``reduced'' Hilbert spaces in their
own right, with the dual Hilbert space being the spaces of the
$\braar{\psi}$. \Eqn{eqn:B3} can then be seen as a
\emph{bi-orthonormality} relation (see \refcite{akhiezer81} and App.
A.8.1 of \refcite{muynk02}) expressing the orthonormality of the
$\ketl{t,\threep}$ basis with respect to this inner product and
allowing for the resolution of the identity
\begin{equation} \label{eqn:B5}
    \intthree p\, (2\Ep)\ketarlz{t,\threep}\bra{\threep} = 1 \,.
\end{equation}
This can be used to reproduce the usual probabilistic interpretation
of quantum mechanics over 3-space for each time $t$ (for further
details, see \refcite{seidewitz06a}).

Further, writing
\begin{equation*}
    \ketarlz{t_{0}, \threep}
        = (2\pi)^{-1/2} \me^{\mp\mi\Ep t_{0}}
                \int \dif p^{0}\, \me^{\mi p^{0}t_{0}} \ketplz \,,
\end{equation*}
where
\begin{equation*}
    \ketplz \equiv (2\pi)^{-2} \intfour x\, \me^{\mi p \cdot x} 
                                            \ketlz{x}
\end{equation*}
is the corresponding 4-momentum state, it is straightforward to see
from \eqn{eqn:B1} that the time limit of \eqn{eqn:B2} is
\begin{equation*}
    \ketarthreep \equiv \lim_{t \to \pm\infty} \ketartp
                     = (2\pi)^{1/2} (2\Ep)^{-1} \ketarEplz \,.
\end{equation*}
Thus, a normal particle ($+$) or antiparticle ($-$) that has
3-momentum $\threep$ as $t \to \pm\infty$ is \emph{on-shell}, with
energy $\pm\Ep$. Such on-shell particles are unambiguously normal
particles or antiparticles, independent of choice of coordinate
system, and, because of the bi-orthonormality relation of
\eqn{eqn:B3}, we can assign classical probabilities for them to have
specific 3-momenta.

\subsection{Fields and Interactions} \label{sect:formalism:fields}

Multiple particle states can be straightforwardly introduced as
members of a Fock space over the Hilbert space of position states
$\ketxl$. First, in order to allow for multiparticle states with
different types of particles, extend the position state of each
individual particle with a \emph{particle type index} $n$, such that
\begin{equation*}
    \inner{x',n';\lambda}{x,n;\lambda}
        = \delta^{n'}_{n}\delta^{4}(x'-x) \,.
\end{equation*}
Then, construct a basis for the Fock space of multiparticle states as
sym\-me\-trized products of $N$ single particle states:
\begin{equation*} 
    \ket{\xnliN}
        \equiv (N!)^{-1/2} \sum_{\text{perms }\Perm}
        \ket{\xni{\Perm 1};\lambda_{\Perm 1}} \cdots
        \ket{\xni{\Perm N};\lambda_{\Perm N}} \,,
\end{equation*}
where the sum is over all permutations $\Perm$ of $1, 2, \ldots, N$.
(Since, for simplicity, I am only considering scalar particles in the
present work, only Bose-Einstein statistics need be accounted for.)

It is then convenient to introduce a \emph{creation field} operator
$\oppsit(x,n;\lambda)$ such that
\begin{equation*}
    \oppsit(x,n;\lambda)\ket{\xnliN} 
        = \ket{x,n,\lambda;\xnliN} \,,
\end{equation*}
with the corresponding annihilation field $\oppsi(x,n;\lambda)$
having the commutation relation
\begin{equation*}
    [\oppsi(x',n';\lambda), \oppsit(x,n;\lambdaz)]
        = \delta^{n'}_{n}\propsym(x'-x;\lambda-\lambdaz) \,.
\end{equation*}
Further, define
\begin{equation*} 
    \oppsi(x,n) \equiv 
        \int_{\lambdaz}^{\infty} \dl\, \oppsi(x,n;\lambda) \,,
\end{equation*}
so that
\begin{equation*}
    [\oppsi(x',n'), \oppsit(x,n;\lambdaz)]
        = \delta^{n'}_{n}\propsym(x'-x) \,.
\end{equation*}

Now, an individual interaction vertex can be considered an event at
which some number of incoming particles are destroyed and some number
of outgoing particles are created. (Note that I am using the
qualifiers ``incoming'' and ``outgoing'' here in the sense of the path
evolution parameter $\lambda$, not time---which means that we are
\emph{not} separately considering particles and antiparticles at this
point.) Such an interaction can be modeled using a \emph{vertex
operator} constructed from the appropriate number of annihilation and
creation operators.

For example, consider the case of an interaction with two incoming
particles, one of type $n_{A}$ and one of type $n_{B}$, and two
outgoing particles of the same types. The vertex operator for this
interaction is
\begin{equation} \label{eqn:C1}
    \opV \equiv g \intfour x\,
        \oppsit(x,n_{A};\lambdaz)\oppsit(x,n_{A};\lambdaz)
        \oppsi(x,n_{A})\oppsi(x,n_{A})
\end{equation}
where the coefficient $g$ represents the relative probability
amplitude of the interaction.

In the following, it will be convenient to use the special adjoint
$\oppsi\dadj$ defined by
\begin{equation*}
    \oppsi\dadj(x,n) = \oppsit(x,n;\lambdaz) \text{ and }
    \oppsi\dadj(x,n;\lambdaz) = \oppsit(x,n) \,.
\end{equation*}
With this notation, the expression for $\opV$ becomes
\begin{equation*}
    \opV = g \intfour x\,
             \oppsi\dadj(x,n_{A})\oppsi\dadj(x,n_{B})
             \oppsi(x,n_{A})\oppsi(x,n_{B}) \,.
\end{equation*}

To account for the possibility of any number of interactions, we just 
need to sum up powers of $\opV$ to obtain the \emph{interaction
operator}
\begin{equation} \label{eqn:C2}
    \opG \equiv \sum_{m=0}^{\infty} \frac{(-\mi)^{m}}{m!}\opV^{m}
              = \me^{-i\opV} \,,
\end{equation}
where the $1/m!$ factor accounts for all possible permutations of the
$m$ identical factors of $\opV$. Note that, unlike the usual
scattering operator, there is no time ordering in the summation here.
(More on this in \sect{sect:decoherence:scattering}.)

The $-\mi$ factors are introduced in \eqn{eqn:C2} so that $\opG$ is
unitary relative to the special adjoint (that is, $\opG\dadj\opG =
\opG\opG\dadj = 1$), so long as $\opV$ is self-adjoint relative to it
(that is, $\opV\dadj = \opV$). The self-adjointness of $\opV$ implies
that an interaction must have the same number of incoming and outgoing
particles, of the same types, at least when only one possible type of
interaction is involved (as is the case with the example of
\eqn{eqn:C1}). The formalism can be easily extended to allow for
multiple types of interactions by adding additional terms to the
definition of $\opV$. In this case, only the overall operator $\opV$
needs to be self-adjoint, not the individual interaction terms.

Now, clearly we can also construct a Fock space from the 3-momentum
representation states $\ketlz{t,\threep}$ and $\ket{t,\threep}$. We
can then define the multiparticle time-limit states
\begin{equation*}
    \begin{split}
        \bra{\threep'_{1\pm},n'_{1};\ldots} 
            &\equiv \lim_{t'_{i} \to \pm\infty} 
               \bra{t'_{1},\threep_{1\pm},n'_{1};\ldots} \,, \\
        \ketlz{\threep_{1\pm},n_{1};\ldots}
            &\equiv \lim_{t_{i} \to \mp\infty}
               \ket{t_{1},\threep_{1\pm},n_{1},\lambdaz;\ldots} \,.
    \end{split}
\end{equation*}
In these states, each particle is \emph{either} a normal particle
($+$) \emph{or} and antiparticle ($-$). Note that the limit is taken
to $+\infty$ for outgoing particles, but to $-\infty$ for outgoing
antiparticles (and vice versa for incoming particles).

These multiparticle 3-momentum states can be used with the interaction
operator $\opG$ to compute multipoint interaction amplitudes. For
example, the four point amplitude for one incoming particle, one
incoming antiparticle, one outgoing particle and one outgoing
antiparticle is given by
\begin{multline} \label{eqn:C4}
    G(\adv{\threepp_{1}{}}, n'_{1}; \ret{\threepp_{2}{}}, n'_{2} |
      \adv{\threep_{1}{}}, n_{1}; \ret{\threep_{2}{}}, n_{2};
      \lambdaz) \\
        = (2\E{\threepp_{1}}2\E{\threepp_{2}}
           2\E{\threep_{1}}2\E{\threep_{2}})^{1/2}
          \bra{\adv{\threepp_{1}{}}, n'_{1}; \ret{\threepp_{2}{}}, n'_{2}}
          \opG
          \ketlz{\adv{\threep_{1}{}}, n_{1}; \ret{\threep_{2}{}}, n_{2}}
          \,.
\end{multline}
(The $2\Ep$ factors are required by the resolution of the identity for
the multiparticle 3-momentum states, generalizing the single particle
case of \eqn{eqn:B3}.) Expanding $\opG$ as in \eqn{eqn:C2} gives a sum
of Feynman diagrams for possible number of interactions. The
time-limited 3-momentum states give the correct truncated amplitudes
for the external legs of the diagrams \cite{seidewitz06a}.

\section{Decoherence} \label{sect:decoherence}

The bi-orthonormality condition of \eqn{eqn:B3} already provides an
example of decoherence. The operator $(2\Ep) \ketarlz{t,\threep}
\bra{\threep}$ represents the quantum proposition that a particle or
antiparticle has a coarse-grained history in which it is free with
3-momentum $\threep$. The fact that these operators are orthogonal by
\eqn{eqn:B3} and resolve the identity by \eqn{eqn:B4} indicates that
these histories are decoherent and classical probabilities can be
assigned as to whether a particle is in one such history or another
\cite{griffiths02}.

In this section I will explore further this concept of decohering
histories of particle paths. I will start with the familiar case of
the two slit experiment, to provide a heuristic example of the
analysis of measurement-induced decoherence using the spacetime path
formalism. This is followed by consideration of scattering experiments
and then, finally, extension of these ideas to the universe as a
whole.

\subsection{Two Slit Experiment} \label{sect:decoherence:2slit}

The canonical two-slit experiment has, of course, been analyzed
several times previously, both in terms of path integrals and
decoherence (see, for example, \refcite{feynman65, hartle91a,
hartle95, halliwell04}). Nevertheless, it is still instructive to use 
this familiar case as a means for introducing the application of the
formalism defined in \sect{sect:formalism}.

Presume that incoming particles are prepared to have a fixed
3-momentum $\threep$. Then, we can take a particle emitted at time
$\tz$ to be in the 3-momentum state $\ketalz{\tz,\threep}$. Further,
assume that the flight time is long enough that, when the
particles reach the slits, they can be considered to be in the
on-shell state $\ketathreep$.

For the purposes of the discussion here, it is sufficient to further
idealize the experiment by considering the slits to be single points
at positions $\threex_{i}$, for $i = 1,2$. The state for the particle
to reach one or the other of the slits is then
\begin{equation*}
    \keta{\threex_{i}{}} = 
        (2\pi)^{-3/2} \intthree p\, 
            e^{-\mi\threep\cdot\threex_{i}} \ketathreep \,.
\end{equation*}
From \eqn{eqn:B3a}, the corresponding probability amplitudes are
\begin{equation*}
    \phi_{i} 
        = \inner{\adv{\threex_{i}{}}}{\tz,\adv{\threep};\lambdaz}(2\Ep) 
        = e^{\mi\threep\cdot\threex_{i}} \,,
\end{equation*}
corresponding to an incoming plane wave. Taking the plane of the slits
to be perpendicular to the direction of $\threep$ results in $\phi_{1}
= \phi_{2} = 1$, corresponding to the equal probability of the
particle reaching any point on that plane. Since the particle is
blocked from passing except through the slits, we can clearly
renormalize the $\phi_{i}$ so that
\begin{equation*}
    \phi_{1} = \phi_{2} = \frac{1}{\sqrt{2}} \,.
\end{equation*}

Suppose the particle passes through the slit at $\threex_{i}$ at some
time $t_{i}$. One can now consider its remaining path separately,
starting at $(t_{i},\threex_{i})$ and ending at some position
$\threex$ on the final screen of the experiment. Qualitatively, the
amplitude for this can be given by
\begin{equation*}
    \psi_{i}(\threex) =
        \inner{\adv{\threex}}{t_{i},\adv{\threex_{i}{}};\lambdaz} \,.
\end{equation*}        
The amplitude for passing through either slit and reaching $\threex$
is then
\begin{equation} \label{eqn:D1}
    \psi(\threex) 
        = \phi_{1}\psi_{1}(\threex) + \phi_{2}\psi_{2}(\threex)
        = \frac{1}{\sqrt{2}}
          ( \inner{\adv{\threex}}{t_{1},\adv{\threex_{1}{}};\lambdaz} + 
            \inner{\adv{\threex}}{t_{2},\adv{\threex_{2}{}};\lambdaz}
          ) \,.
\end{equation}

The result of the experiment is a measurement made of the final
position $\threex$. This measurement is represented by a measuring
instrument eigenstate $\ket{m(\threex)}$ such that
\begin{equation} \label{eqn:D2}
    \inner{m(\threex')}{m(\threex)} = 
        \delta^{3}(\threex' - \threex) \,.
\end{equation}
The measurement eigenstate $\ket{m(\threex)}$ must be weighted by the
amplitude $\psi(\threex)$ for the particle to reach position
$\threex$. From the point of view of particle paths, each state
$\psi(\threex)\ket{m(\threex)}$ can be viewed as representing the
entire coarse-grained history of a particle being emitted, passing
through one or the other of the slits and being measured as arriving
at position $\threex$. Due to the orthogonality condition of
\eqn{eqn:D2}, these coarse-grained history states do not interfere
with each other---that is, the histories \emph{decohere}, so a
classical probability of $\sqr{\psi(\threex)}$ can be assigned to
them. From \eqn{eqn:D1}, it is clear that this probability will,
however, include interference effects between the slit-specific
amplitudes $\psi_{1}$ and $\psi_{2}$.

We can, of course, also represent the less-coarse-grained histories
for the particle passing through \emph{just} one slit as
$\psi_{i}(\threex) \ket{m(\threex)}$, for $i = 1,2$. But these
histories do \emph{not} decohere, since
\begin{equation*}
    \psi_{1}^{*}(\threex)\psi_{2}(\threex)
        \inner{m(\threex)}{m(\threex)}
\end{equation*}
is not zero. (Actually, with the delta function normalization of
\eqn{eqn:D2} this value is infinite, but that would not be so for a
more realistic instrument with finite resolution.)

Suppose, however, that we add a measuring device that measures whether
the particle passes through slit 1 or slit 2. This device has two
eigenstates denoted $\ket{s(i)}$, for $i = 1,2$, such that
\begin{equation*}
    \inner{s(i)}{s(j)} = \delta_{ij} \,.
\end{equation*}
The coarse-grained history for a particle being measured as passing
through slit $i$ then being measured as reaching position $\threex$ is
now $\psi_{i}(\threex)\ket{s(i)}\ket{m(\threex)}$. These histories now
\emph{do} decohere, since
\begin{equation*}
    \psi_{i}^{*}(\threex')\psi_{j}(\threex)
        \inner{s(i)}{s(j)}\inner{m(\threex')}{m(\threex)} =
        \sqr{\psi_{i}(\threex)} \delta_{ij} 
        \delta^{3}(\threex' - \threex) \,,
\end{equation*}
and they can be given the individual probabilities
$\sqr{\psi_{i}(\threex)}$.

The results of this analysis are, of course, as would be expected.
Notice, however, that, rather than the usual approach of time evolving
states, the approach here constructs states representing entire
coarse-grained particle histories. Measurements are modeled as being
coupled to specific points in these histories. Thus, rather than
modeling some initial state of a measuring instrument evolving into a 
state with a specific measurement, the states $\ket{s(i)}$ and
$\ket{m(\threex)}$ represent the occurrence of specific measurement
values \emph{as part of} the overall history of the experiment.

The occurrence of a specific measurement value places a constraint on
the possible particle paths that can be included in any coarse-grained
history consistent with that measurement. Thus, $\ket{s(i)}$ places
the constraint that paths must pass through slit $i$, while
$\ket{m(\threex)}$ places the constraint that the paths end at
position $\threex$. If a coarse-grained history includes \emph{all}
possible paths consistent with the constraints for specific
measurement values, and no others, then the orthogonality of the
measurement states causes such a history to decohere from other
similar histories for different measurement values.

In this sense, the tensor product of the measurement eigenstates
provides a complete, orthogonal basis for decoherent coarse-grained
histories of the experiment. Given observations of certain measurement
values, the experiment, as a whole, can be said with certainty to be
``in'' the specific history eigenstate selected by those measurement
values. Nothing definitive, however, can be said about finer-grained
histories, since these histories do not decohere.

The important point here is that the experiment is not modeled as
``evolving'' into a decoherent state. Rather it is \emph{entire
coarse-grained histories} of the experiment that decohere, with
observed measurements simply identifying which actual history was
observed.

\subsection{Scattering} \label{sect:decoherence:scattering}

We now turn to the more general problem of multiparticle scattering,
with the goal of providing an analysis similar to that provided for
the two slit experiment in \sect{sect:decoherence:2slit}. Clearly, we 
can base this on the multiple particle interaction formalism discussed
in \sect{sect:formalism:fields}.

However, the formulation of \eqn{eqn:C4} is still not that of the
usual scattering matrix, since the incoming state involves particles
at $t \to +\infty$ but antiparticles at $t \to -\infty$, and vice
versa for the outgoing state. To construct the usual scattering
matrix, it is necessary to have incoming multiparticle states that are
composed of individual asymptotic particle states that are all
consistently for $t \to -\infty$ and outgoing states with individual
asymptotic states all for $t \to +\infty$. That is, we need to shift
to considering ``incoming'' and ``outgoing'' in the sense of
\emph{time}.

To do this, we can take the viewpoint of considering antiparticles to
be positive energy particles traveling forwards in time, rather than
negative energy particles traveling backwards in time. Since both
particles and their antiparticles will then have positive energy, it
becomes necessary to explicitly label antiparticles with separate
(though related) types from their corresponding particles. Let $\na$
denote the type label for a normal particle type and $\nr$ denote the
corresponding antiparticle type.

For normal particles of type $\na$, position states are defined as
in \eqn{eqn:B0}:
\begin{equation*}
    \inner{x,\na}{\xz,\na;\lambdaz} = \thetaax\prop \,.
\end{equation*}
For antiparticles of type $\nr$, however, position states are now
defined such that
\begin{equation*} 
    \inner{x,\nr}{\xz,\nr;\lambdaz} = \thetaax\propsym(\xz - x) \,.
\end{equation*}
Note the reversal with respect to \eqn{eqn:B0a} of $\xz$ and $x$ on
the right side of this equation.

Carrying through the derivation for antiparticle 3-momentum states
based on the new antiparticle states $\ket{x,\nr}$ does, indeed, give
positive energy states, but with reversed three momentum
\cite{seidewitz06a}:
\begin{equation*}
    \ket{t,\threep,\nr}
        = (2\Ep)^{-1}\int_{-\infty}^{t} \dt_{0}\,
          \ketlz{\tz,\threep,\nr} \,,
\end{equation*}
where
\begin{equation*}
    \ketlz{t_{0},\threep,\nr} = \ketlz{\adv{t_{0},-\threep},n} \,.
\end{equation*}
Further, taking the limit $t \to +\infty$ gives the on-shell states
\begin{equation*}
    \ket{\threep,\nr}
        \equiv \lim_{t \to +\infty} \ket{t,\threep,\nr}
             = (2\pi)^{1/2}(2\Ep)^{-1}\ketlz{+\Ep,-\threep} \,.
\end{equation*}

We can now reasonably construct Fock spaces with single time,
multiparticle basis states
\begin{equation*}
    \ketlz{t;\pnariN}
        \equiv \ketlz{t,\threep_{1},n_{1\pm};\ldots;
                      t,\threep_{N},n_{N\pm}} \,,
\end{equation*}
over all combinations of particle and antiparticle types and,
similarly,
\begin{equation*}
    \ket{t;\pnariN} 
        \equiv \ket{t,\threep_{1},n_{1\pm};\ldots;
                    t,\threep_{N},n_{N\pm}} \,.
\end{equation*}
We can then take consistent time limits for particles and antiparticles
alike to get the incoming and outgoing states
\begin{equation*}
    \begin{split}
        \ketlz{\pnariN} &= \lim_{t \to -\infty}\ketlz{t;\pnariN} \,, \\
        \ket{\pnariN} &= \lim_{t \to +\infty}\ket{t;\pnariN} \,.
    \end{split}
\end{equation*}

Reorganizing the interaction amplitude of \eqn{eqn:C4} in terms of
these new asymptotic states gives the more usual form using the
scattering operator $\opS$. Showing explicitly the asymptotic time
limit used for each particle:
\begin{equation} \label{eqn:E1}
    \begin{split}
        \bra{+\infty, \adv{\threepp_{1}{}}, n'_{1}; 
            &-\infty, \ret{\threepp_{2}{}}, n'_{2}}
         \opG
         \ketlz{-\infty, \adv{\threep_{1}{}}, n_{1}; 
                +\infty, \ret{\threep_{2}{}}, n_{2}} \\
            &= \bra{+\infty, \threepp_{1}, \adv{n'_{1}{}}; 
                    +\infty, \threep_{2}, \ret{n_{2}{}}}
               \opS
               \ketlz{-\infty, \threep_{1}, \adv{n_{1}{}}; 
                      -\infty, \threepp_{2}, \ret{n'_{2}{}}} \\
            &= \bra{\threepp_{1}, \adv{n'_{1}{}}; 
                    \threep_{2}, \ret{n_{2}{}}}
               \opS
               \ketlz{\threep_{1}, \adv{n_{1}{}}; 
                      \threepp_{2}, \ret{n'_{2}{}}} \,.
    \end{split}
\end{equation}

More generally, consider applying $\opS$ to an incoming state of $N$
particles, giving $\opS\ket{\pnarlziN}$. Using the resolution of the
identity
\begin{multline} \label{eqn:E2}
    \sum_{N = 0}^{\infty}\, \sum_{\advret{n_{i}{}}}
    \int \dthree p_{1} \cdots \dthree p_{N}\,
        \left[ \prod_{i=1}^{N} 2\E{\threep_{i}} \right] \\
        \times \ket{\pnarlziN}\bra{\pnariN}
        = 1 \,,
\end{multline}
expand the state $\opS\ket{\pnarlziN}$ as
\begin{multline*}
    \opS\ket{\pnarlziN} \\
        = \sum_{N' = 0}^{\infty}\, \sum_{\advret{n_{i}{}}}
          \int \dthree p'_{1} \cdots \dthree p'_{N'}\,
          \left[ \prod_{i=1}^{N'} 2\E{\threepp_{i}} \right]
          \ket{\pnparlziN} \\
          \times \bra{\pnpariN}\opS\ket{\pnarlziN} \,.
\end{multline*}
This shows how $\opS\ket{\pnarlziN}$ is a superposition of possible
out states, with the square of the scattering amplitude giving the
probability of a particular out state for a particular in state.

Note that each operator 
\begin{equation*}
    \ket{\pnarlziN}\bra{\pnariN}
\end{equation*}
represents not the proposition that the particles have the 3-momenta
$\threep_{i}$ at any one point in time, but, rather, that they have
these momenta \emph{for their entire history}. Since, by \eqn{eqn:E2},
these operators orthogonally resolve the identity, these histories do
not interfere with each other and are thus trivially decoherent. This
is why the square of the scattering amplitude gives a classical
probability.

It should also be noted that both $\ket{\pnarlziN}$ and
$\opS\ket{\pnarlziN}$ represent states of the entire ``universe''
under consideration. The state $\ket{\pnarlziN}$ represents a universe
in which all particles remain free and there are no interactions. This
free particle state does not evolve into $\opS\ket{\pnarlziN}$.
Rather, $\opS\ket{\pnarlziN}$ is the state of a \emph{different}
universe, in which interactions \emph{do} occur. The operator $\opS$
simply provides a convenient method for constructing the states of the
interacting particle universe from the states of the free particle
universe.

\subsection{Probabilities} \label{sect:decoherence:probabilities}

The decoherence of coarse-grained histories allows for a 
mathematically consistent assignment of probabilities. Physically, 
the concept of ``probability'' here is to be interpreted as meaning 
the likelihood that an arbitrary selection from the population of all 
possible coarse-grained histories will yield a specific history. In 
other words, the greater the probability assigned to a history, the 
more likely it is that it is actually the history of the ``universe'' 
under consideration.

Of course, it is not immediately clear how the assignment of 
probabilities to entire histories relates to the statistics of 
physical results of measurement processes occuring within those 
histories. Before continuing, I would like to briefly 
consider this point.

To simplify further discussion, let a single Greek letter, say
$\alpha$, represent an entire configuration $\threep_{1},
\threep_{2}, \ldots$ of on-shell particle 3-momenta. In this
notation, incoming states $\ket{\pnarlziN}$ are denoted as simply
$\ketlz{\alpha}$ and outgoing states $\ket{\pnpariN}$ become
$\ket{\alpha'}$. The resolution of the identity from \eqn{eqn:E2} is
then
\begin{equation*}
    \int \dif\alpha\, \ketlz{\alpha}\bra{\alpha} = 1 \,,
\end{equation*}
where $\int \dif\alpha$ denotes the entire set of integrals and
summations. 

Suppose the same scattering experiment is repeated, independently, $n$
times. Let $\ketlz{\psi_{i}}$ be the asymptotic free incoming state
for the $i$-th repetition. Considered all together, the overall free
particle state of this ``universe'' of experiments is
\begin{equation*}
    \ketlz{\psi} = \ketlz{\psi_{1}}\cdots\ketlz{\psi_{n}} \,.
\end{equation*}
The state $\opS\ketlz{\psi}$ is then the superposition of all possible 
histories of interactions among the incoming particles. At a large 
enough time after all the experiments take place, the outgoing 
particles should be on-shell in a state $\bra{\alpha} = 
\bra{\alpha_{1},\ldots,\alpha_{n}}$, where each $\bra{\alpha_{i}}$ is 
the outgoing state for the $i$-th repetition, and the probability for 
this overall result is $\sqr{\bra{\alpha}\opS\ketlz{\psi}}$.

If we can neglect interactions between each experiment repetition,
then the scattering amplitude should approximately factor:
\begin{equation*}
    \bra{\alpha}\opS\ketlz{\psi}
        \approx \bra{\alpha_{1}}\opS\ketlz{\psi_{1}} \cdots
                \bra{\alpha_{n}}\opS\ketlz{\psi_{n}} \,.
\end{equation*}
(If the repetitions are widely spacelike separated, then this follows
from the cluster decomposition of $\opS$ \cite{weinberg95,
horwitz81}.) Thus, the overall probability for scattering into
$\alpha$ is approximately the product of the scattering probabilities
for each cluster. 

Now, consider a measurement $m(\alpha_{i})$ taken of each experimental
result. Suppose the measurement determines in which member of a
disjoint partition of values $\alpha_{i}$ lies. The probability
amplitude for a measurement of $\alpha_{i}$ to have the specific
(discrete) value $m_{i}$ is
\begin{equation*}
    \psi_{i}(m_{i}) \equiv
        \int_{m_{i}} \dif \alpha_{i}\,
            \bra{\alpha_{i}}\opS\ketlz{\psi_{i}} \,,
\end{equation*}
where the integration is over the subset of values corresponding to
the measurement result $m_{i}$. Assuming identical preparation for the
experiments, the $\psi_{i}$ should all be the same function $\psi(m)$.

The overall weighted measurement state is then
\begin{equation} \label{eqn:F1}
    \psi(m_{1}) \cdots \psi(m_{n})
        \ket{m_{1}} \cdots \ket{m_{n}} \,,
\end{equation}
where $\ket{m_{i}}$ is the measuring instrument eigenstate for the
measurement of the $i$-th experimental result. Once again, this
overall state represents a specific coarse-grained history in which
the specific measurement results $m_{1}, \ldots, m_{n}$ are obtained
for the $n$ repetitions of the scattering experiment. The question to
be asked is how the relative frequency of any given result in this set
compares to the quantum mechanically predicted probabilities
$\sqr{\psi(m_{i})}$ (see also \refcites{hartle68, graham73} for 
discussions of this question in the context of traditional and 
many-worlds interpretations of quantum mechanics).

Define the \emph{relative frequency} of some specific measurement
result $\ell$ within the set $m_{1}, \ldots, m_{n}$ to be
\begin{equation} \label{eqn:F1a}
    f_{\ell}(m_{1}, \ldots, m_{n}) \equiv
        \frac{1}{n} \sum_{i = 1}^{n} \delta_{m_{i}\ell} \,.
\end{equation}
Since this relative frequency is itself an observable, a relative
frequency operator $\op{F}_{\ell}$ can be defined which has relative
frequencies as its eigenvalues:
\begin{equation*}
    \op{F}_{\ell}\ket{m_{1}} \cdots \ket{m_{n}} =
        f_{\ell}(m_{1}, \ldots, m_{n}) 
        \ket{m_{1}} \cdots \ket{m_{n}} \,.
\end{equation*}
Define the average
\begin{equation*}
    \avg{\op{F}_{\ell}}
        \equiv \sum_{m_{1} \ldots m_{n}}
                    f_{\ell}(m_{1}, \ldots, m_{n})
                    \sqr{\psi(m_{1})} \cdots
                    \sqr{\psi(m_{n})} \,.
\end{equation*}
Substituting \eqn{eqn:F1a} and using the normalization
$\sum\sqr{\psi(m_{i})} = 1$ then gives \cite{hartle68, graham73}
\begin{equation} \label{eqn:F2}
    \avg{\op{F}_{\ell}} = \sqr{\psi(\ell)} \,.
\end{equation}

\Eqn{eqn:F2} is mathematically consistent with the probability
interpretation of quantum mechanics. However, this mathematical
average still needs to be connected to physical results. To do this,
consider a further measurement, this time of the relative frequency
$\op{F}_{\ell}$. Note that this is a measurement \emph{of the previous
measurements} $m_{i}$, perhaps simply by counting the records of the
results of those measurements. The new measurement results are thus
the functions $f_{\ell}(m_{1}, \ldots, m_{n})$, with corresponding
eigenstates $\ket{f_{\ell}(m_{1}, \ldots, m_{n})}$.

The overall state
\begin{equation} \label{eqn:F3}
    \psi(m_{1}) \cdots \psi(m_{n})
        \ket{m_{1}} \cdots \ket{m_{n}}
        \ket{f_{\ell}(m_{1}, \ldots, m_{n})}
\end{equation}
then represents the history in which a specific relative frequency is
measured for a specific set of scattering results. Since these history
states are still decoherent due to the original set of measurement
states, the total probability for observing a certain relative
frequency $f_{\ell}$ is given by the sum of the probabilities for each
of the states for which $nf_{\ell}$ of the $m_{i}$ have the value
$\ell$. This probability is
\begin{equation*}
    p(f_{\ell}) = \binom{n}{nf_{\ell}} p_{\ell}^{nf_{\ell}}
                      (1-p_{\ell})^{n(1-f_{\ell})} \,,
\end{equation*}
where $p_{\ell} = \sqr{\psi(\ell)}$.

The probability $p(f_{\ell})$ is just a Bernoulli distribution. By the
de Moivre--Laplace theorem, for large $n$, this distribution is
sharply peaked about the mean $f_{\ell} = p_{\ell} =
\avg{\op{F}_{\ell}}$. Thus, the probability becomes almost certain
that a choice of one of the histories \eqref{eqn:F3} will be a history
in which the observed relative frequency will be near the prediction
given by the usual Born probability interpretation. Of course, for
finite $n$, there is still the possibility of a ``maverick'' universe
in which $f_{\ell}$ is arbitrarily far from the expected value---but
it would seem that (in most cases, at least) our universe is simply
not one of these.

There have been a number of criticisms in the literature of using
relative frequency as above as the basis for the quantum probability
interpetation (see, for example, \refcites{kent90, squires90}).
However, these criticisms relate to attempts to actually justify the
Born probability interpretation itself. My goal here is more modest:
to simply show that, assuming the Born probability rule applies for
history states, that the statistics of repeated measurement results
within such a history would be expected to follow a similar rule. In
this regard, criticisms of, e.g., circularity and the need for
additional assumptions, do not apply here. (For justification of the
Born rule itself for quantum states, the arguments of Zurek based on
``environment-assisted invariance'' \cite{zurek03, zurek05} would seem
to be relevent, but I will not pursue this further here.)

\subsection{Cosmological States} \label{sect:decoherence:cosmo}

Extending the ideas from \sect{sect:decoherence:probabilities}, let
$\ketlz{\Psi}$ be the \emph{cosmological state} representing the
free-particle evolution of the universe from the initial condition of
the big bang. Then $\opS\ketlz{\Psi}$ is a superposition of all
possible interacting particle histories of the universe. Obviously,
this really should also include interactions leading to bound states,
not just scattering. For the purposes of the present discussion,
however, it is sufficient to simply allow that some of the products of
the scattering interactions may be composite particles rather than
fundamental.

A specific coarse-grained history in this superposition can be
identified by a specific configuration $\alpha$ of all classically
observable particles throughout the life of the universe. (For the
present discussion, assume that this is a large but finite number of
particles.) In this case, $\Psi(\alpha) =
\bra{\alpha}\opS\ketlz{\Psi}$ might reasonably be called the ``wave
function of the universe'', since $\sqr{\Psi(\alpha)}$ is the
probability of the universe having the configuration $\alpha$ given
its cosmological state $\opS\ketlz{\Psi}$. (Clearly, for this to be
the true wave function of the universe, $\opS$ would need to include
the effects of all the actual types of interactions, including gravity
\cite{hartle83}.) Further, given that the universe can be decomposed
into approximately isolated subsystems, the overall probability
$\sqr{\Psi(\alpha)}$ will approximately factor into a product of
probabilities for the histories of each of the subsystems.

Now, consider that any classically measurable quantity should be a
function of some subset of the classical configuration $\alpha$.
Divide $\alpha$ into $\alpha_{1}, \alpha_{2}, \ldots$ (this division
need not be complete or disjoint), and let $m_{i}(\alpha_{i})$
represent the result of a measurement made on the subset $\alpha_{i}$.
We can then represent a measuring instrument for $m_{i}$ as having a
set of orthogonal states $\ket{m_{i}(\alpha_{i})}$ representing the
various possible measurement outcomes.

Of course, a measuring instrument is, itself, a part of the universe
being measured. And a complete theory of measurement would have to
account for how such an instrument, as a subsystem of the universe,
becomes correlated with some other part of the universe and itself
decoheres into non-interfering states. However, it is not the intent
of this paper to present such a complete theory. (For a discussion of 
related issues in a non-relativistic context, see \refcites{zurek98, 
zurek03} and the references given there.)

For our purposes here, it is sufficient to consider a ``measurement
process'' to be a process that produces a persistent record of
distinguishable results correlated with the measured subsystem, based
on classical variables. By definition, such a process can be
abstracted into a representation by orthogonal result states. We can
then extend the kind of analysis used in \sect{sect:decoherence:2slit}
for the two slit experiment, and consider the complete measurement
state of the universe to be
\begin{equation} \label{eqn:G1}
    \Psi(\alpha_{1}, \alpha_{2}, \ldots) 
        \ket{m_{1}(\alpha_{1})}\ket{m_{2}(\alpha_{2})} \ldots \,,
\end{equation}
in which the measurement results are 
correlated with the corresponding configuration of the universe with 
probability amplitude given by the wave function $\Psi(\alpha_{1}, 
\alpha_{2}, \ldots)$.

Further, suppose some of the measurements are of relative frequencies 
of results of repeated experiments. Then, by extension of the 
argument in \sect{sect:decoherence:probabilities}, for a large enough 
number of repetitions within a ``typical'' history, the observed 
relative frequency will accurately reflect the probabilities as 
predicted by quantum theory.

It is worth emphasizing again that the universe does not ``evolve
into'' the state \eqref{eqn:G1}. Rather, this state represents a
\emph{complete} coarse-grained history of the universe, in which the
measurement values $m_{1}(\alpha_{1}), m_{2}(\alpha_{2}), \ldots$ are
observed, implying the corresponding classical configuration
$\alpha_{1}, \alpha_{2}, \ldots$ for the universe. The correlation of
the measurements with the configuration of the universe means that the
measurement results effectively provide information on which
coarse-grained history the universe is ``really in.''

It is in exactly this sense that the universe can be represented as
the eigenstate \eqref{eqn:G1} of the measurements made within it.

\section{Conclusion} \label{sect:conclusion}

I would like to conclude with some remarks on the interpretational 
implications of the concept of cosmological states defined in 
\sect{sect:decoherence:cosmo}.

Each cosmological state $\ket{\alpha_{1}, \alpha_{2}, \ldots}$, with
corresponding measurement state \eqref{eqn:G1}, represents a possible,
complete, coarse-grained history of the universe. Of course, each such
course-grained history is still a quantum superposition of many
fine-grained histories. However, if we include in the $m_{i}$ all the
measurements made in the entire history of the universe, then the
corresponding measurement states are the finest-grained possible that
can be determined by inhabitants of the universe.

The measurement states themselves are decoherent and orthogonal, but
the distribution of measurement results in any specific coarse-grained
history will still show the effects of interference of the superposed
fine-grained histories (as we saw in the simple case of the two slit
experiment in \sect{sect:decoherence:2slit}). This reflects the fact
that such interference effects really are observed in our universe.

Now, all measurements ever made so far determine only some very small
portion of a configuration $\alpha$ of the universe. Nevertheless, in
principle, it is consistent to consider all such measurements to be,
indeed, made on a portion of some overall $\alpha$, selecting a
specific classical history from the family given by
$\opS\ketlz{\Psi}$, and that this is the ``real'' history of the
universe. The formalism here allows for no further judgement on the
``real'' history of the universe beyond the coarse-grained
superpositions determined by the measurement results.

This conception is very much in the spirit of the original work by
Everett \cite{everett57} on what has become known as the ``many
worlds'' interpretation. The key point is that there is no need to
consider any sort of observation by observation ``collapse of the wave
function.'' Rather, consistent measurement results are determined by
appropriately decohering histories \cite{hartle95, halliwell03,
halliwell04}, and known measurement results constrain the possible
histories.

However, Everett and his successors \cite{dewitt73} generally
considered the dynamic evolution of states in time. In this
formulation, a measurement process at a certain time causes a state to
``branch'' into orthogonal components, one for each possible
measurement result. This leads almost inevitably to the conception of
the continual dynamic creation of ``many worlds,'' only one of which
is ever really apparent to any observer.

In contrast, in the approach presented here, entire coarse-grained
histories of the universe decohere for all time. It is only necessary
to consider one of these to be the ``real'' history of the actual
universe, though we have only very partial information on which
history this actually is. There is no need to consider the other
histories to have any ``real'' existence at all. Nevertheless, within
the ``real'' history of our universe, all observations made at the
classical level will be distributed according to the probabilistic
rules of quantum theory.

Instead of a ``no collapse'' interpretation, this is, in a sense, a
``one collapse'' interpretation---the single collapse of the wave
function of the universe into the cosmological state of the entire
coarse-grained history of the universe. It is as if God did indeed
play dice with the universe, but that He threw very many dice just
once, determining the fate of the universe for all space and time.

\bibliography{../../RQMbib}

\end{document}